\documentclass[10pt, conference]{IEEEtran}
\usepackage{amssymb}
\usepackage{pstricks}
\usepackage{amsmath}
\usepackage[dvips]{graphicx}
\usepackage{enumerate}
\usepackage{ifthen}

\newtheorem{theorem}{Theorem}

\newtheorem{lemma}{Lemma}

\IEEEoverridecommandlockouts

\newcommand{\Ind}[1]{
1_{\left\{#1\right\}}
}

\title{Uplink Performance of Large Optimum-Combining Antenna Arrays in Power-Controlled Cellular Networks}
\author{\IEEEauthorblockN{Siddhartan Govindasamy $^\dagger$}
\IEEEauthorblockA{F. W. Olin College of Engineering. Needham, MA,  USA\\
Email: siddhartan.govindasamy@olin.edu}
\thanks{\noindent $ ^\dagger$ Sponsored in part by the National Science Foundation (CCF-111721)}}

\begin{document}

\maketitle
\pagenumbering{arabic}

\begin{abstract}
The uplink of  interference-limited cellular  networks with base stations that have large numbers of antennas and use linear Minimum-Mean-Square Error (MMSE) processing with power control is analyzed. Simple approximations, which are exact in an asymptotic sense, are provided for the spectral efficiencies (b/s/Hz) of links in these systems. It is also found that when the number of base-station antennas is moderately large, and the number of mobiles in the entire network is large,  correlations between the transmit powers of mobiles within a given cell do not significantly influence the spectral efficiency of the system. As a result, mobiles can perform simple power control (e.g. fractional power control) that does not depend on other users in the network, reducing system complexity and improving the analytical tractability of such systems.
\end{abstract}


\section{Introduction}

%\subsection{Background and Main Contributions}
Massive Multiple-Input Multiple-Output (MIMO) has been proposed as a technology to meet the exponential growth in demand for wireless communications that is forecasted for the near future. These systems use  large antenna arrays at base stations to separate signals to/from users  spatially, enabling simultaneous, co-channel transmissions (see e.g. \cite{marzetta2010noncooperative, LarssonMag}). While it is known that the spatial matched filter performs optimally when the number of base station antennas is large, it is also known that there is a significant range of realistic system parameters for which the linear MMSE receiver can provide spectral efficiencies equivalent to systems with matched-filter receivers with approximately an order of magnitude more antennas \cite{hoydis2013massive}. Hence, it is interesting to study the performance of systems with the MMSE receiver and moderately large numbers of base-station antennas.

The uplink  of massive MIMO systems with MMSE processing has been analyzed before in \cite{hoydis2013massive}, \cite{YatesMMSEMassive}, and \cite{wangsinr}. However, the spatial distribution of users and base stations was not considered in these works. Explicitly modeling spatial user and base station distributions can provide insight into the effects of tangible system parameters such as the density of base stations and users, and the number of antennas, on data rates. 

  Modeling the spatial distribution of mobiles and base stations for uplink channels is known to be challenging as noted in \cite{novlan2012analytical}. In that work, the complexity associated with modeling the spatial distribution of mobiles and base stations is handled by forming Voronoi cells about the mobiles and distributing one base station per Voronoi cell.  This approach results in the transmitters in the network being distributed  according to a Poisson Point Process (PPP) for which a number of analytical techniques have been developed recently. In other works that consider power control, which can be used to manage interference as well as conserve battery life,  Monte Carlo simulations are used (e.g. \cite{novlan2012analytical, rao2007reverse}). In \cite{coupechoux2011set}, the authors proposed a method for analyzing power control on the uplink of cellular networks which is based on approximating the interference as originating from sources in concentric rings about a test base station. These approximations that have been taken in the literature underscores the difficulty in analyzing the uplink of power-controlled cellular sytems. 

In systems with many antennas at  base stations however, asymptotic effects can be utilized to address the complexity of the uplink, without having to make significant approximations on the spatial user and base-station distributions.  In \cite{CellularNetworks}, we considered the uplink of cellular networks with a similar model as that used here, except we considered a specific power control model in which mobiles attempt to invert the effects of the path loss between themselves and their respective base stations. Since the locations of the mobiles were assumed to be independent in that work, the transmit powers of the mobiles were also independent, simplifying analysis. In another related work \cite{govindasamy2014uplink}, we considered the uplink of a cellular network with large antenna arrays at Poisson-distributed base stations, but with a limited number of active mobiles per cell and no power control.

In this work, we consider a cellular network with hexagonal cells and multiantenna linear MMSE receivers. We assume a  general power control algorithm where mobiles are assigned transmit powers as a function of the positions of all mobiles in their respective cells, relative to their base stations. Thus, the transmit powers of the mobiles in each cell are correlated. Note that this correlation increases the complexity of the analysis as compared to the system with independent transmit powers we considered in \cite{CellularNetworks}. Further discussion of this is provided in Section III-B. We find that when the number of antennas at the base station is moderately large and the network is interference limited, with a large number of mobiles, the correlation between the transmit powers of mobiles within each cell does not signficantly impact the achievable data rates. Hence, the transmit power of each mobile does not need to depend on other mobiles in its cell. This finding has significant practical implications as it simplifies power control computation and reduces the need for base-stations to assign transmit powers to mobiles in amplitude-reciprocal systems as described further in Section III-B.

\section{System Model}\label{Sec:NetworkModel}

Consider a 2-dimensional  network with base stations located on a hexagonal grid and where  mobiles connect to their closest base stations. Let the density of base stations be $\rho_c = 1/A_c$, where $A_c$ is the area of each cell. We further assume that there is a base station at the origin, with the locations of the base stations denoted by $B_0, B_1, B_2, \cdots$. $B_0 = o$  shall be referred to as the representative base station. Let the cells be denoted by $\bar{\mathcal{C}}_0, \bar{\mathcal{C}}_1, \cdots$. Suppose that there is a mobile at  $X_0$, which we call the representative mobile, located at some point in the cell containing the origin.   In the remainder of this work, we shall analyze  the link between the representative transmitter and the representative base station which we shall also refer to as the representative link.
\begin{figure}
\center
\includegraphics[width = 2.75in]{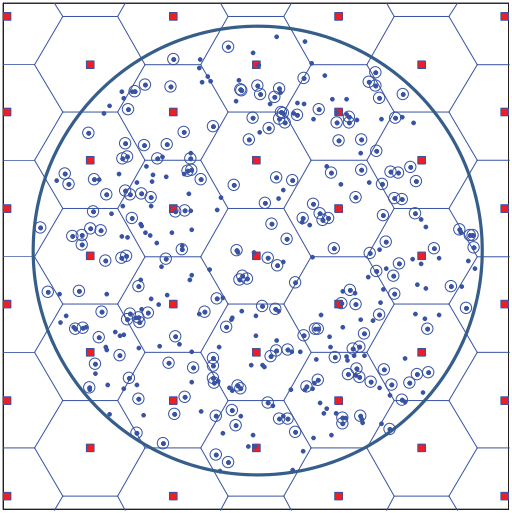}
\caption{Illustration of a cellular network with at most $K = 10$ users transmitting with non-zero powers. Mobiles are represented by the dots and circles are used to highlight the active mobiles.}
\label{Fig:Cells}
\end{figure}

Overlaid on the network of base stations is a radius-$R$ circular network centered at the origin as shown in Figure \ref{Fig:Cells}. In this network, $n$ additional mobiles are  at  independent and identically distributed (i.i.d.) locations.  These mobiles, if active will be co-channel interferers to the representative link. Let   $X_1, X_2, \cdots, X_{n}$ represent the positions of the mobiles,  and $r_i = |X_i|$ be the distance of the $i$-th mobile from the origin. The density of potentially active mobiles
$\rho_m$ satisfies
\vspace{0.3cm}
\begin{align}
n = \pi\, \rho_m \, R^2. \label{Eqn:MaternPotentialInterferers}
\end{align}

\vspace{0.3cm}
We denote the cell in which the $i$ th mobile falls as $\mathcal{C}_i$. The average power (averaged over the fast fading) received at each antenna of the representative base station from a mobile  transmitting with power $P_i$, from a distance $r_i$, is $P_ir_i^{-\alpha}$. The path-loss exponent $\alpha > 2$ is assumed to be a rational number. The transmit power of the $i$-th mobile is a function of the location of all  mobiles in its cell, relative to the base station in that cell. In other words 
\begin{align}
P_i = g\left(\left\{X_j-\mbox{Center}(\mathcal{C}_i):X_j\in\mathcal{C}_i\right\}\right) \leq P_M\,, \label{Eqn:PowerControlDef}
\end{align}
where $\mbox{Center}(\mathcal{C}_i)$ is the location of the base station at the center of the cell $\mathcal{C}_i$. Moreover, we require that the mobiles either be assigned zero transmit power or a positive transmit power that is bounded from below by $P_{\ell b}$. In other words, if $P_i > 0$, then $P_i > P_{\ell \, b}$. This is a technical requirement that is reasonable in practical systems as mobiles can be expected to have a minimum power if they are transmitting. 

\eqref{Eqn:PowerControlDef} describes a general power control algorithm where the transmit power of a mobile is a function of the positions of all the mobiles in its  cell relative to the location of the base station. Additionally, we restrict  the total number of mobiles transmitting in each cell to $K$ by requring that for each $i$, 
\begin{align}
\left|\left\{j: X_j \in \bar{\mathcal{C}}_i  \mbox{ and } P_j > 0 \right\}\right| \leq K.
\end{align}
This restriction helps to meet quality of service requirements and to permit channel estimation using orthogonal pilot sequences as has been proposed for massive MIMO systems in the literature (e.g. see \cite{LarssonMag}). 
%If a cell has less than or equal to $K$ mobiles, they are all active and if it has greater than $K$ mobiles, $K$ of them are selected to transmit at random, and with uniform probaility. This ensures that at most $K\geq 1$ mobiles are active in each cell.  
We assume that the representative transmitter is always active, i.e. $P_0 > 0$. At a given sampling time, the sampled signals at the $N$ antennas of the representative base station are represented in  $\mathbf{y}\in\mathbb{C}^{N\times 1}$ as follows.
\begin{align}
\mathbf{y} = r_0^{-\frac{\alpha}{2}} \mathbf{g}_0 \sqrt{P_0} x_0 + \sum_{i= 1}^n r_i^{-\frac{\alpha}{2}} \mathbf{g}_i \sqrt{P_i} x_i\,, \label{Eqn:SysEq}
\end{align}
where  $x_i$, is the  transmitted symbol of  the $i$-th mobile and $\mathbf{g}_i \in \mathbb{C}^{N\times 1}$ contains i.i.d., zero-mean,
unit  variance, circularly symmetric, complex Gaussian random variables denoted by $\mathcal{CN}(0,1)$. Since we focus on the interference-limited regime where the noise power $\to$ 0,  \eqref{Eqn:SysEq} does not include noise, which means that our results are applicable to networks with a high density of mobiles.

The asymptotic regime considered here is the limit as $N$, $n$ and $R \to \infty$, such that $n/N = c$ and $\rho_m$ are constant, and \eqref{Eqn:MaternPotentialInterferers} holds. For the remainder of this paper,  whenever  $n,N$ or $R \to \infty$, it is assumed that the other two quantities go to infinity as well.   Note that while our analysis may provide insight into the scaling behavior of such systems, we use it as a tool to analyze large networks of a \emph{fixed} size.  We additionally require $\lim_{n\to\infty}\Pr(P_i > 0) n > N$  which  ensures that as $R\to\infty$, with high probability, there would be a larger number of actively transmitting mobiles in the \emph{entire} network than antennas at the representative base station. This ensures that the interference covariance  matrix $\mathbf{R}$ defined below is invertible with probability 1 for all sufficently large $n, N, R$.
\begin{align}
\mathbf{R} = \sum_{j = 1}^n r_j^{-\alpha}{P_j} \mathbf{g}_{j} \mathbf{g}_j^\dagger \label{Eqn:InterfCovDef}
\end{align}
 We assume that $\mathbf{R}$ and $\mathbf{g}_0$ are known at the representative base station. Thus, the results here can be interpreted as bounds on the performance of real systems which will be subject to channel estimation errors which may be significant in certain cases when the number of antennas is very large.

%The main results are given in terms of limiting values of a normalized version of the SIR, $\beta_N = N^{-\alpha/2}\,P_0,|X_0|^\alpha\,\mbox{SIR}$, at the output of the MMSE receiver.% which estimates $x_0$ using the weight vector
%$\mathbf{w}$.  The estimate, $\hat{x}_0 = \mathbf{w}^\dagger \mathbf{y}$, with
%\begin{align}
%&\mathbf{w}^\dagger = \nu\, \mathbf{g}_0^\dagger\left(\sum_{j = 1}^n r_j^{-\alp%ha}{P_j} \mathbf{g}_{j} \mathbf{g}_j^\dagger\right)^{-1}\,, \label{Eqn:PCWeight%}
%\end{align}
%where $\nu$ is a scale factor that does not impact the SIR.   % Note that $\mathbf{R}$ can be estimated using blind techniques and could be acquired with relatively low overhead.
%However, it should be noted that an accurate estimate of $\mathbf{R}$ requires a number of samples on the order of the number of receiver antennas,
%thus the coherence time of the system must be sufficiently long to permit accurate estimation of this quantity.
%Conditioned on $P_0$, the SIR at the output of the receiver  is given by
%\begin{align}
%  \text{SIR} = \frac{P_0\,r_0^{-\alpha}\left|\mathbf{w}^\dagger \mathbf{g}_0\ri%ght|^2}{\sum_{i = 1}^n r_i^{-\alpha}P_i\left|\mathbf{w}^\dagger \mathbf{g}_i\right|^2}\,. \label{Eqn:GenSIR}
%\end{align}

\section{Asymptotic Spectral Efficiency}
\subsection{Main Results}

Define a normalized version of the SIR at the output of the MMSE receiver,
\begin{align}
\beta_N = N^{-\frac{\alpha}{2}} \frac{r_0^\alpha}{P_0} \mbox{SIR}.
\end{align}
Using the standard formula for the SIR at the output of the MMSE receiver, we can write
\begin{align}
\beta_N  = N^{-\frac{\alpha}{2}}  \mathbf{g}_0^\dagger \mathbf{R}^{-1} \mathbf{g}_0\,. \label{Eqn:NormSIRDef}
\end{align}

With these definitions, conditioned  $r_0$, $P_0$, and denoting the probability-density function (PDF) of transmit powers by $f_P(p)$,  we have the following theorem.
\begin{theorem} \label{Theorem:TetheredSIR}
Consider the network model from Section \ref{Sec:NetworkModel}.
As  $N, n, R \to \infty$, $\beta_N\to \beta$ in probability where $\beta$ is the
non-negative solution to the following equation
\begin{align}
&E[P^{{\frac{2}{\alpha}}}]\beta^{{\frac{2}{\alpha}}} \left[\frac{\pi}{\alpha} \csc \left({\frac{2\pi}{\alpha}}\right)\right] - \frac{2\pi\rho_m \beta}{\alpha }\int_{0}^{P_M \left(\frac{\pi\rho_m}{c}\right)^{\frac{\alpha}{2}}}\frac{\tau^{-\frac{2}{\alpha}}}{1+\tau \beta}  \times  \nonumber \\
&\int_{\tau \left(\frac{c}{\pi\rho_m}\right)^{\frac{\alpha}{2}}}^\infty f_P(x)x^{\frac{2}{\alpha}} dx\, d\tau = \frac{1}{2\rho_m\pi}  \label{Eqn:TheoremTetheredSINR}
\end{align}
\end{theorem}
{\it Proof:} Given in Appendix \ref{Sec:MainProof}.

Moreover, since $P_i < P_M$,  we can bound the second term on the LHS of \eqref{Eqn:TheoremTetheredSINR} as
\begin{align}
&\frac{2\pi\rho_m \beta}{\alpha }\int_{0}^{P_M \left(\frac{\pi\rho_m}{c}\right)^{\frac{\alpha}{2}}}\frac{\tau^{-\frac{2}{\alpha}}}{1+\tau \beta} \int_{\tau \left(\frac{c}{\pi\rho_m}\right)^{\frac{\alpha}{2}}}^\infty f_P(x)x^{\frac{2}{\alpha}} dx\, d\tau\leq\nonumber\\
&\frac{2\pi\rho_m \beta}{\alpha }\int_{0}^{P_M \left(\frac{\pi\rho_m}{c}\right)^{\frac{\alpha}{2}}}\frac{\tau^{-\frac{2}{\alpha}}}{1+\tau \beta} P_M^{\frac{2}{\alpha}} d\tau\,.
\end{align}

If $c\to\infty$ after $n,N,R\to\infty$, the integral converges to zero. Hence, we have
\begin{align}
\lim_{c\to\infty}\lim_{N\to\infty} \beta_N = \left[\frac{ \alpha}{2\pi^2\rho_m E\left[P^{\frac{2}{\alpha}}\right]}\sin\left(\frac{2\pi}{\alpha}\right)\right]^{\frac{\alpha}{2}}
\end{align}
Additionally, if we assume all mobiles transmit using Gaussian codebooks, by the continuous-mapping theorem (e.g.  \cite{bliss2013adaptive} (3.103)) and removing the normalization of the SIR
\begin{align}
&\lim_{c\to\infty}\lim_{N\to\infty}\left|\log_2\left(1+\mbox{SIR}\right)- \right.\nonumber\\
& \left.\log_2\left(1+P_0r_0^{-\alpha}\left[\frac{N\alpha}{2\pi^2\rho_m E\left[P^{\frac{2}{\alpha}}\right]}\sin\left(\frac{2\pi}{\alpha}\right)\right]^{\frac{\alpha}{2}}\right)\right| = 0\,\label{Eqn:SpecEff}
\end{align}
in probability. Further, by applying the bounded-convergence theorem mirroring steps in Appendix E in \cite{TxCSIJournal}, we find that
\begin{align}
%&\lim_{c\to\infty}\lim_{N\to\infty}\left|E\left[\log_2\left(1+\mbox{SIR}\right)%\right]-\right.\nonumber\\
%& \left.\log_2\left(1+P_0r_0^{-\alpha}\left[\frac{N \alpha}{2\pi^2\rho_m E\left%[P^{\frac{2}{\alpha}}\right]}\sin\left(\frac{2\pi}{\alpha}\right)\right]^{\frac%{\alpha}{2}}\right)\right| = 0\,
&\lim_{c\to\infty}\lim_{N\to\infty}\left|E\left[\log_2\left(1+\mbox{SIR}\right)\right]-\log_2\left(1+\mbox{SIR}\right)\right|\to 0.
\end{align}
Hence, for systems with large antenna arrays at each base station but a much larger number of mobiles in the \emph{entire} network than the number of antennas, the spectral efficiency $\gamma$ and its mean are well approximated as follows
\begin{align}
&E[\gamma] \approx \gamma \approx\nonumber \\
&\log_2\left(1+P_0r_0^{-\alpha}\left[\frac{N\alpha}{2\pi^2\rho_m E\left[P^{\frac{2}{\alpha}}\right]}\sin\left(\frac{2\pi}{\alpha}\right)\right]^{\frac{\alpha}{2}}\right) \label{Eqn:SpecEff}
\end{align}
If we now assume that $P_0$ is a random variable,  for large $N,n,R$, but with $n\gg N$, we can approximate the CDF of the spectral efficiency as
\begin{align}
  F_\gamma(\gamma) \approx  F_q\left(\left(2^{{\gamma}}-1\right)\left[\frac{N\alpha}{2\pi^2\rho_m E\left[P^{\frac{2}{\alpha}}\right]}\sin\left(\frac{2\pi}{\alpha}\right)\right]^{-\frac{\alpha}{2}}\right)\label{Eqn:HexCellSpecEffCDF}
\end{align}
where $F_q(q)$ is the CDF of $q = P_0r_0^{-\alpha}$.

\subsection{Discussion}
Note that the spectral efficiency is primarily a function of the transmit power of the representative mobile and $ E\left[P^{\frac{2}{\alpha}}\right]$ which means that there is little dependence between the transmit powers of mobiles within a cell which greatly simplifies the power control computation. Moreover, in systems which use full-duplex operation and where uplink and downlink channels have amplitude reciprocity, mobiles can estimate the path losses between themselves and their base stations and control their transmit powers, without having the knowledge of the transmit powers or path-losses of the other mobiles in their cell. Thus, mobiles can control their transmit power without requiring the base station to communicate information to them which results in a reduction in the protocol overhead.

Finally, it is worth noting that Theorem 1 in this work has a very similar form to Theorem 1 in \cite{CellularNetworks}. However, the assumptions used here are quite different which requires a different proof. In both these works, the main result is a consequence of the convergence in probability of the empirical distribution function (e.d.f.) of $N^{\frac{\alpha}{2}}P_i r_i^{-\alpha}$, for $i = 1, 2, \cdots, n$. In \cite{CellularNetworks}, it was assumed that the transmit power of a given mobile was solely a function of the distance between itself and its given base station. Thus, the transmit powers of the mobiles were not correlated. As a result, the e.d.f. converges as a direct consequence of the law large numbers.  In this work, the correlation between transmit powers results in a more complicated proof of the convergence of the e.d.f. of  $N^{\frac{\alpha}{2}}P_i r_i^{-\alpha}$, for $i = 1, 2, \cdots, n$ as given in Appendix \ref{Sec:MainProof}.

\section{Numerical Simulations and Results} \label{Sec:Numerical}

We conducted Monte Carlo simulations to verify the accuracy of our asymptotic findings. In all cases, we simulated 1000 realizations of a cellular network according to the system model, with path-loss exponent $\alpha = 4$, density of mobiles $\rho_m = 1$ and the density of base stations $\rho_c = 10^{-4}$. In all cases, the number of mobiles assigned non-zero transmit power $K$ is fixed. Thus, we use a high relative density of mobiles to base-stations to ensure that with high probability, each cell will have $K$ actively transmitting mobiles (i.e. mobiles assigned non-zero transmit powers).

\begin{figure}
\center
\includegraphics[width = 3.5in]{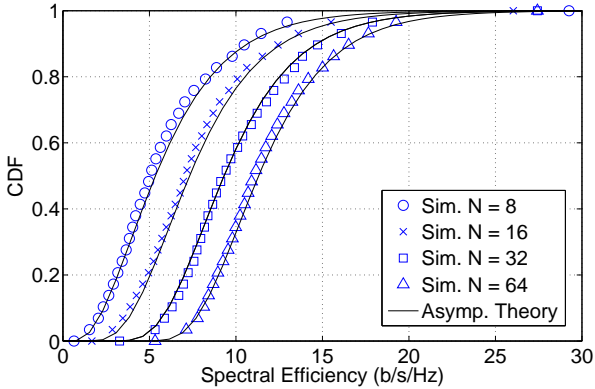}
\caption{Simulated and asymptotic CDFs of spectral efficiency for N = 8,16, 32  and 64 antennas, and $K = 10$ with correlated transmit powers. Mobiles closer to their base station transmit with power 0.5 and mobiles farther away  from their base station transmit with unit power as described in Section  \ref{Sec:Numerical}.
}
\label{Fig:CDF2}
\end{figure}

In Figure \ref{Fig:CDF2} we show simulated spectral efficiencies of an ad hoc power control algorithm
which serves mainly to illustrate that correlation between transmit powers does not influence the spectral efficiency significantly. Mobiles are assigned three different transmit powers, $P_\ell = 0.5$, $P_h = 1$ and zero. First, at most $K$ mobiles from each cell are selected to transmit with equal probability from the mobiles in each cell. In each cell, half of the mobiles  closest to the base station are assigned power $P_\ell$ and the other half are assigned $P_h$.   Using order statistics and the PDF of $r_0$  which is available in \cite{CellularNetworks}, we can  compute the CDF of the spectral efficiency using \eqref{Eqn:HexCellSpecEffCDF}. This computation is straightforward but tedious and is not included here for the sake of breivity. The resulting CDF is illustrated by the solid lines in the figure and the simulated values are represented by the markers. The close agreement between the theoretical prediction and     simulations verifies the accuracy of  \eqref{Eqn:HexCellSpecEffCDF} when there is a high degree of correlation between transmit powers in a given cell.

\begin{figure}
\center
\includegraphics[width = 3.5in]{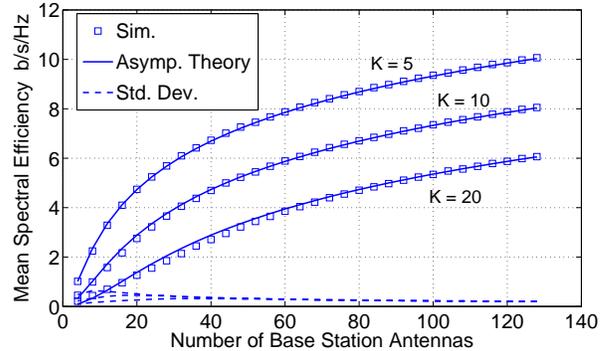}
\caption{Simulated mean spectral efficiency with path-loss inversion power control. Markers represent simulated results, solid lines represent the asymptotic predictions and the dashed lines are simulated standard deviations of the spectral efficiency. The maximum number of actively transmitting mobiles in each cell is limited to the values of $K$ indicated in the plot.}
\label{Fig:MeanFracPC}
\end{figure}

\begin{figure}
\center
\includegraphics[width = 3.75in]{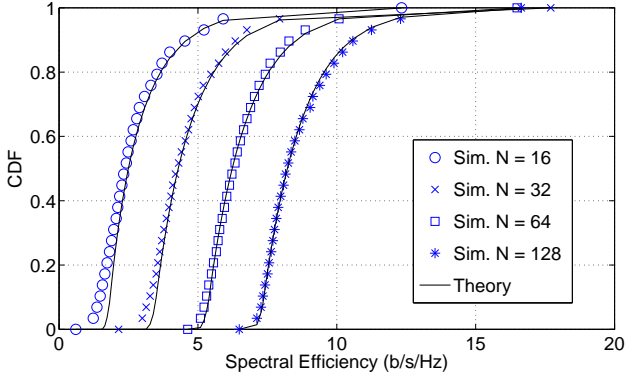}
\caption{Simulated and theoretical CDF of the spectral efficiency with fractional power control, where the $i$-th mobile transmits with power $|X_i-\mathcal{C}_i|^{\alpha\, \epsilon}$. The markers represent simulated values and the solid lines represent the asymptotic approximation. The maximum number of actively transmitting mobiles per cell is $K = 10$.}
\label{Fig:MeanFracPC2}
\end{figure}

Figures \ref{Fig:MeanFracPC} and \ref{Fig:MeanFracPC2} illustrate systems which use fractional power control (see e.g. \cite{novlan2012analytical}). As described in the previous paragraph at most $K$ mobiles  transmit with non-zero power per cell, and any mobile which is less than unit distance from its base station is assigned zero power in order to avoid  unbounded received signal power at base stations. Of the mobiles selected to transmit,  a mobile at $X_i$ transmits with power $|X_i - \mbox{Center}(\mathcal{C}_i)|^{\alpha\, \epsilon}$, where $\epsilon \in [0,1]$. When $\epsilon = 0$, the mobiles transmit with equal power and when $\epsilon = 1$, mobiles invert the effects of the path losses between themselves and their respective base stations.  

 Figure \ref{Fig:MeanFracPC} illustrates  the mean spectral efficiency vs. number of antennas when $\epsilon = 1$. Note that since the path-loss is inverted in this model, the spectral efficiency of a given link is approximately equal to \eqref{Eqn:SpecEff}. The term  $E[P_i^{\frac{2}{\alpha}}]$ for this model can be obtained by elementary calculus using existing results from \cite{CellularNetworks} and is not included here for brevity. Note here that the simulated mean spectral efficiencies and the asymptotic prediction are extremely close which helps verify the accuracy of the asymptotic results. Moreover, the standard deviation (and variance) of the spectral efficiency decays with $N$ which supports the conclusion that the spectral efficiency approaches its asymptote in probability since mean square convergence implies convergence in probability.

\begin{figure}
\center
\includegraphics[width = 3.75in]{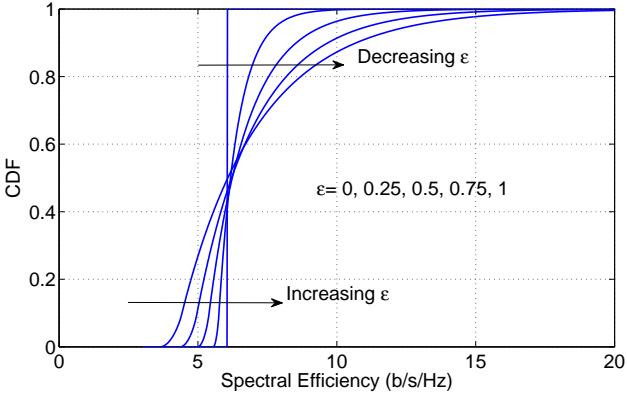}
\caption{CDF of the spectral efficiency based on the asymptotic approximation for $\epsilon = 0, 0.25, 0.5, 0.75, 1$ and $N = 64$. Observe that for low outage probabilities, $\epsilon = 1$ results in the largest spectral efficiency.}
\label{Fig:MeanFracPCMultE}
\end{figure}

 Figure \ref{Fig:MeanFracPC2} illustrates simulated CDFs of the spectral efficiency and its asymtotic approximation from \eqref{Eqn:HexCellSpecEffCDF}, with $\epsilon = 0.5$.  $F_q(q)$ for this model can be computed using straightforward, but tedious calculations by starting from the CDF of the distance of a random point in a hexagonal cell to the center of the hexagon, given in \cite{CellularNetworks}. Observe that with increasing $N$, the accuracy of the asymptotic predictions improves, becoming virtually indistinguishable  from the simulations            when the number of antennas is $N = 128$.  The asymptotic approximation to the CDF of the spectral efficiency for $\epsilon = 0, 0.25, 0.5, 0.75, 1$ and $N = 64$ is plotted in Figure \ref{Fig:MeanFracPCMultE}. This figure indicates that for low outage probabilities, $\epsilon  = 1$, i.e. path-loss inversion is the optimal strategy for fractional power control.

\section{Summary and Conclusions}
We analyze the uplink spectral efficiency in spatially distributed cellular networks with large numbers of base-station antennas and          power control. Base stations are assumed to use the linear MMSE receiver, and transmit powers of mobiles within the same cell are correlated. Simple asymptotic approximations, which are precise in an asymptotic sense, are provided for the spectral efficiency of a representative link in such networks. It is also found that when the number of base-station antennas is moderately large, and the number of mobiles in the entire network is very large, the correlation between the transmit powers of mobiles in the same cell does not effect the spectral efficiency significantly. As a result, simple power control algorithms with no dependency between transmit powers of mobiles is sufficient in such networks. This finding thus helps reduce the complexity of practical systems and simplfies their analysis. In future work, optimal power allocation strategies based on this framework shall be explored. 

\section{Acknowledgement}
We thank Dr. Gunnar Peters of Huawei Research for suggesting this work, and the reviewers for helpful comments.

\appendix
\subsection{Proof of Main Result} \label{Sec:MainProof}
The main result is proved using Lemma 1 of \cite{CorrTransPaper} which is repeated here for convenience.

\begin{lemma} \label{Lemma:SIRConvergenceLemma}
Consider the quantity
\begin{align}
\gamma_N = \frac1N \mathbf{s}^\dagger\left(\frac1N\mathbf{S \Psi S}^\dagger\right)^{-1}\mathbf{s} \label{eqn20}
\end{align}
where $\mathbf{s} \in \mathbb{C}^{N\times 1}$ and $\mathbf{S}
\in \mathbb{C}^{N\times n}$ comprise i.i.d., zero-mean,
unit-variance entries, $n/N = c$, and  $\mathbf{\Psi} =
\mbox{diag}(\psi_1, \psi_2, \cdots \psi_{n})$. Suppose that as
$n, N \to\infty$, the e.d.f. of $(\psi_1, \psi_2, \cdots \psi_{n})$ converges in probability to  $H(x)$.
Additionally, assume that there exists an $n_0$ such that
$\forall n > n_0$, $\lambda_{min}\left(\frac1N\mathbf{S \Psi S}^\dagger\right) \geq \lambda_{\ell b} > 0$, with probability 1, where $\lambda_{min}(\mathbf{A})$  denotes the minimum eigenvalue of any square matrix $\mathbf{A}$, and $\lambda_{\ell b}$ is an arbitrary, strictly positive number. Then,
$\gamma_N \to \gamma$ in probability where $\gamma$ equals
the unique non-negative real solution for $\gamma$ in
\begin{equation}
1 =
\gamma\,c\int_{0}^{\infty} \frac{\tau dH(\tau)}{1+\tau \gamma}\,.\label{InitialFixedPoint}
\end{equation}
\end{lemma}
Comparing \eqref{eqn20} to \eqref{Eqn:NormSIRDef} and \eqref{Eqn:InterfCovDef}, we observe that 
\begin{align}
\beta_N = r_0^{-\alpha}\, P_0 \gamma  
\end{align}
if $\mathbf{s} = \mathbf{g}_0$, the columns of $\mathbf{S}$ are $\mathbf{g}_1, \mathbf{g}_2, \cdots \mathbf{g}_n$ and $\mathbf{\Psi} = diag\left(N^{\alpha/2} P_1\, r_1 ^{-\alpha}, N^{\alpha/2} P_2\, r_2 ^{-\alpha}, \cdots, N^{\alpha/2} P_n\, r_n ^{-\alpha} \right)$. Thus to use Lemma \ref{Lemma:SIRConvergenceLemma}, we need to show that the e.d.f. of $N^{\alpha/2} P_1\, r_1 ^{-\alpha}, N^{\alpha/2} P_2\, r_2 ^{-\alpha}, \cdots, N^{\alpha/2} P_n\, r_n ^{-\alpha}$ converges in probability to a limiting function and find its form. Additionally, we need to show that the minimum eigenvalue property required in  Lemma \ref{Lemma:SIRConvergenceLemma} is satisfied by this model. 

Let $H_n(x)$ denote the empirical distribution function of   $N^{\alpha/2} P_1\, r_1 ^{-\alpha}, N^{\alpha/2} P_2\, r_2 ^{-\alpha}, \cdots, N^{\alpha/2} P_n\, r_n ^{-\alpha}$.
The first condition is proved in the following lemma.
\begin{lemma}\label{Lemma:EDFConvergenceLemma}
As $n,N, R\to\infty$, $H_n(x)$ converges  in probability to
\begin{align}
&H(x) = 1-\frac{\rho_m\, x^{-\frac{2}{\alpha}}}{c}E[P_j^{\frac{2}{\alpha}}]\nonumber \\
 &-\int_{ x\left(\frac{c}{\pi\rho_m}\right)^{\frac{\alpha}{2}}}^{P_M} \left(1-x^{-\frac{2}{\alpha}}\frac{\rho_m P_j^{\frac{2}{\alpha}}}{c}\right)f_P(P_j)dP_j.\label{Eqn:LimCDFLemma}
\end{align}
\end{lemma}
 {\it Proof: Given in Appendix \ref{Sec:ProofofEDFConvergenceLemma}}

The minimum eigenvalue condition is satisfied as shown in Appendix \ref{Sec:MinEigenvalueProof}. Note that \eqref{Eqn:LimCDFLemma} has the same form as the corresponding equation in  \cite{CellularNetworks}, even though  it was assumed in \cite{CellularNetworks} that transmit power of the mobiles are independently distributed. Following the steps used to prove Theorem 1 in \cite{CellularNetworks} and the fact that $f_P(P_i) = 0$ for $P_i > P_M$ , we find that $\beta_N\to\beta$ in probability, where $\beta$ is the unique positive real solution to \eqref{Eqn:TheoremTetheredSINR}.
%\begin{align}
%&E[P^{{\frac{2}{\alpha}}}]\beta^{{\frac{2}{\alpha}}} \left[\frac{\pi}{\alpha} \csc \left({\frac{2\pi}{\alpha}}\right)\right] \nonumber \\
%&- \frac{2\pi\rho_w \beta}{\alpha }\int_{0}^{\infty}\frac{\tau^{-\frac{2}{\alpha}}}{1+\tau \beta}  \int_{\tau \left(\frac{c}{\pi\rho_w}\right)^{\frac{\alpha}{2}}}^\infty f_P(x)x^{\frac{2}{\alpha}} dx\, d\tau = \frac{1}{2\rho_w\pi}\.
%\end{align}

%\begin{align}
% &H(x) = F_P\left( x\left(\frac{c}{\pi\rho_m}\right)^{\frac{\alpha}{2}}\right)+%\frac{\rho\, x^{-\frac{2}{\alpha}}}{c}E[P_j^{\frac{2}{\alpha}}]\nonumber\\
%&\;\;\;\;\;\;\;\;\;\;\;\;\;\;\;\;\;\;\;\;\;\;\;\;-x^{-\frac{2}{\alpha}}\int_{ x\left(\frac{c}{\pi\rho_m}\right)^{\frac{\alpha}{2}}}^{P_M} \frac{\rho P_j^{\frac{2}{\alpha}}}{c}f_P(P_j)dP_j.\label{Eqn:LimCDFLemma2}
%\end{align}
%\end{lemma}
%Taking the derivative of $H(x)$ with respect to $x$ and simplifying yields
%\begin{align}
%\frac{dH(x)}{dx} = \frac{2\pi\rho_w}{c\alpha}E&\left[P^{\frac{2}{\alpha}}\right%]x^{-\frac{2}{\alpha}-1}  \nonumber \\
%&-  \frac{2\pi\rho_w}{c\alpha}x^{-\frac{2}{\alpha}-1}\int_{x\left(\frac{c}{\pi\rho_m}\right)^{\frac{\alpha}{2}}}^\infty f_P(\tau)\tau^{\frac{2}{\alpha}}d\tau.  \label{Lemma:DiffCorrelatedPowers}
%\end{align}
%Note that \eqref{Eqn:LimCDFLemma2} 

\subsection{Proof of Lemma \ref{Lemma:EDFConvergenceLemma}}\label{Sec:ProofofEDFConvergenceLemma}
 Let $\bar{H}_n(x) = E[H_n(x)]$. Following steps used to derive (38) in \cite{CorrTransPaper}, we have that for each $\epsilon>0$,
\begin{align}
%\begin{split}
& \Pr\left(\left|H_n(x) - \bar{H}_n(x)\right| \geq \frac{\epsilon}{2} \right)\leq \nonumber \\
&\frac{4}{\epsilon^2 n^2} \sum_{i=1}^n \sum_{j=1}^n
\left({\rm{Pr}}\left( P_i\, r_i ^{-\alpha} \leq  x \,N^{-\alpha/2}, P_j\, r_j^{-\alpha},\leq x\,N^{-\alpha/2}\right) \right.\nonumber \\
&\left.-{\rm{Pr}}\left( P_i\, r_i^{-\alpha}\leq x\,N^{-\frac{\alpha}{2}}\right)
{\rm Pr}\left( P_j\, r_j^{-\alpha}\leq x\,N^{-\frac{\alpha}{2}}\right)\right).\label{eqn32}
%\end{split}
\end{align}
Let $R_c$ denote the radius of the smallest circle containing a given cell in its interior, and $\mathcal{C}_A$ denote the union of all cells wholly contained in $B(0,R)$, i.e., $\mathcal{C}_A = \bigcup_{k:\, \bar{\mathcal{C}}_k \subset B(0,R)}\bar{\mathcal{C}}_k$, where $B(Y,D)$ denotes a disk of radius $D$ centered on $Y$. $X_i$ and $X_j$ are in different cells if $|X_i - X_j| > 2R_c$. Let the event that $X_i$ and $X_j$ are in different cells, and $X_i, X_j \in \mathcal{C}_A$ be denoted by $\mathcal{A}$. Note that as $R\to\infty, \rm{Pr}(\mathcal{A})\to 1$ as the probability of $X_i$ and $X_j$ falling in a cell that is not wholly contained in the circular network goes to zero, and the probability that $X_i$ and $X_j$ are separated by a distance less than $2R_c$ goes to zero as $\Theta(R^{-2})$ from  \cite{Mathai}
%Consider the following:
%\begin{align}
%&{\rm{Pr}}\left( \left. P_i\, r_i ^{-\alpha} > x \,N^{-\alpha/2}, \, P_j\, r_j^%{-\alpha},> x\,N^{-\alpha/2}\right|\mathcal{A}, P_i, P_j\right)=\nonumber\\
%&{\rm{Pr}}\left(\left.  r_i  <  \left(\frac{N^{\alpha/2}P_i}{x}\right)^{\frac{1%}{\alpha}}\right| r_j<\left(\frac{N^{\alpha/2}P_j}{x}\right)^{\frac{1}{\alpha}}%,\mathcal{A}, P_i, P_j\right)\times\nonumber\\
%&\;\;\;\;\;\;\;\;\;{\rm{Pr}}\left(\left.  r_j  <  \left(\frac{N^{\alpha/2}P_j}{%x}\right)^{\frac{1}{\alpha}}\right|\mathcal{A}, P_i, P_j\right)\,.
%\end{align}

Note that $P_i$ may be dependent on $|X_i - \mbox{Center}\left(\mathcal{C}_i\right)|$, where $\mbox{Center}\left(\mathcal{C}_i\right)$ denotes the center of the cell $\mathcal{C}_i$, where recall that $\mathcal{C}_i$ is the cell containing $X_i$. Hence, $P_i$ and $r_i = |X_i|$ are  correlated in general. However, we note that since $P_i$ is only a function of the mobiles in $\mathcal{C}_i$, for $R$ sufficiently large, 
\begin{align}
& \Pr(r_i < x|P_i) \leq \Pr(y < x+2R_c) \label{Eqn:CDFDepUBound}
\end{align}
where $y$ is a random variable distributed with uniform probability in $\mathcal{C}_A$. 
%$because it is guaranteed that $|X_i -  \mbox{Center}\left(\mathcal{C}_i\right)|$ < 2R_c$.
Recall that $\mathcal{A}$ is the event that $X_i$ and $X_j$ are in different cells which are wholly contained in $B(0,R)$. For $R$ large enough such that $|\mathcal{C}_A| > A_c$ we can write an upper bound which is based on  \eqref{Eqn:CDFDepUBound} and assuming that $\mathcal{C}_j$ is located outside $B\left(0,\left(\frac{N^{\alpha/2}P_i}{x}\right)^{\frac{1}{\alpha}}\right)$. 
\begin{align}
&{\rm{Pr}}\left(\left.  r_i  <  \left(\frac{N^{\alpha/2}P_i}{x}\right)^{\frac{1}{\alpha}}\right| r_j<\left(\frac{N^{\alpha/2}P_j}{x}\right)^{\frac{1}{\alpha}},\mathcal{A}, P_i, P_j\right)\leq \nonumber\\
&\frac{\pi\left(\left(\frac{N^{\frac{\alpha}{2}}P_i}{x}\right)^{\frac{1}{\alpha}}+2R_c\right)^2}{|\mathcal{C}_A|-A_c}\Ind{0\leq\left(\left(\frac{N^{\frac{\alpha}{2}}P_i}{x}\right)^{\frac{1}{\alpha}}+2R_c\right)^2\leq R^2} \nonumber \\
&\;\;\;\;\;\;\;\;+ \Ind{\left(\left(\frac{N^{\frac{\alpha}{2}}P_i}{x}\right)^{\frac{1}{\alpha}}+2R_c\right)^2> R^2}\label{Eqn:InitUB}
\end{align}
Observing that $P_i$ and $P_j$ are indepeendent since $X_i$ and $X_j$ are in different cells, we can take the expectation of the above expression with respect to $P_i$ as follows.
\begin{align}
&{\rm{Pr}}\left(\left.  r_i  <  \left(\frac{N^{\alpha/2}P_i}{x}\right)^{\frac{1}{\alpha}}\right| r_j<\left(\frac{N^{\alpha/2}P_j}{x}\right)^{\frac{1}{\alpha}},\mathcal{A}\right)\leq \nonumber\\
%\end{align}
%\begin{align}
&E_P\left[\pi\frac{\left(\left(\frac{N^{\frac{\alpha}{2}}P_i}{x}\right)^{\frac{1}{\alpha}}+2R_c\right)^2}{|\mathcal{C}_A|-A_c}\Ind{0\leq\left(\left(\frac{N^{\frac{\alpha}{2}}P_i}{x}\right)^{\frac{1}{\alpha}}+2R_c\right)^2\leq R^2}\right. \nonumber \\
&\left. \;\;\;\;\;\;\;\;+ \Ind{\left(\left(\frac{N^{\frac{\alpha}{2}}P_i}{x}\right)^{\frac{1}{\alpha}}+2R_c\right)^2> R^2}\right]
\end{align}
Next, we take the limit as $N, n, R\to\infty$ in the manner of Theorem 1, and interchange the order of the expectation and limit by the bounded convergence theorem. Noting that as $R\to\infty$, edge effects disappear and $\frac{|\mathcal{C}_A|}{ \pi R^2} \to 1$, we have
\begin{align}
&\lim_{R\to\infty}{\rm{Pr}}\left(\left.  r_i  <  \left(\frac{N^{\frac{\alpha}{2}}P_i}{x}\right)^{\frac{1}{\alpha}}\right| r_j<\left(\frac{N^{\frac{\alpha}{2}}P_j}{x}\right)^{\frac{1}{\alpha}},\mathcal{A}\right)\leq \nonumber\\
&E_P\left[\frac{\pi \rho_m}{c}\left(\frac{P_i}{x}\right)^{\frac{2}{\alpha}}\Ind{0\leq\left(\frac{P_i}{x}\right)^{\frac{2}{\alpha}}\leq \frac{c}{\pi\rho_m}}+ \Ind{ \frac{c}{\pi\rho_m}<\left(\frac{P_i}{x}\right)^{\frac{2}{\alpha}}}\right]\nonumber\\
%\end{align}
%\begin{align}
&=\int_{0}^{ x\left(\frac{c}{\pi\rho_m}\right)^{\frac{\alpha}{2}}}\frac{\pi \rho_m}{c}\left(\frac{P_i}{x}\right)^{\frac{2}{\alpha}}f_P(P_i) dP_i\nonumber \\
&\;\;\;\;\;\;\;\;\;\;\;\;\;\;\;\;\;\;\;\;\;\;\;\;\;\;\;\;\;\;\;\;\;\;\;\;+ \int_{ x\left(\frac{c}{\pi\rho_m}\right)^{\frac{\alpha}{2}}}^{\infty} f_P(P_i)dP_i\nonumber\\
&=\int_{0}^{\infty}\frac{\pi \rho_m}{c}\left(\frac{P_i}{x}\right)^{\frac{2}{\alpha}}f_P(P_i) dP_i\nonumber \\
&\;\;\;\;\;\;\;\;\;\;\;\;\;\;\;+ \int_{ x\left(\frac{c}{\pi\rho_m}\right)^{\frac{\alpha}{2}}}^{\infty} \left(1-\frac{\pi \rho_m}{c}\left(\frac{P_i}{x}\right)^{\frac{2}{\alpha}}\right)f_P(P_i)dP_i\nonumber\\
&=\frac{\pi \rho_m}{c}x^{-\frac{2}{\alpha}}E[P_i^{\frac{2}{\alpha}}]\nonumber \\
&\;\;\;\;\;\;\;\;\;\;+ \int_{ x\left(\frac{c}{\pi\rho_m}\right)^{\frac{\alpha}{2}}}^{\infty} \left(1-\frac{\pi \rho_m}{c}\left(\frac{P_i}{x}\right)^{\frac{2}{\alpha}}\right)f_P(P_i)dP_i\label{Eqn:CDFUBLast}
\end{align}
Similar to \eqref{Eqn:InitUB}, we can write a lower bound as follows.
\begin{align}
&{\rm{Pr}}\left(\left.  r_i  <  \left(\frac{N^{\alpha/2}P_i}{x}\right)^{\frac{1}{\alpha}}\right| r_j<\left(\frac{N^{\alpha/2}P_j}{x}\right)^{\frac{1}{\alpha}},\mathcal{A}, P_i, P_j\right)> \nonumber\\
&\frac{\pi\left(\left(\frac{N^{\frac{\alpha}{2}}P_i}{x}\right)^{\frac{1}{\alpha}}-2R_c\right)^2-A_c}{|\mathcal{C}_A|}\times\nonumber \\
&\;\;\;\;\;\;\;\;\Ind{0\leq\left(\left(\frac{N^{\frac{\alpha}{2}}P_i}{x}\right)^{\frac{1}{\alpha}}-2R_c\right)^2-A_c\leq R^2} \nonumber \\
&\;\;\;\;\;\;\;\;\;\;\;\;\;\;+ \Ind{\left(\left(\frac{N^{\frac{\alpha}{2}}P_i}{x}\right)^{\frac{1}{\alpha}}-2R_c\right)^2-A_c> R^2}\label{Eqn:InitLB}
\end{align}
Comparing \eqref{Eqn:InitLB} and \eqref{Eqn:InitUB}, we note that in the limit as $R\to\infty$, both the upper and lower bounds will co-incide. Thus, \eqref{Eqn:CDFUBLast} holds with equality. Moreover, following a similar set of steps,
we can find that
\begin{align}
&\lim_{R\to\infty}{\rm{Pr}}\left(\left.  r_j  <  \left(\frac{N^{\frac{\alpha}{2}}P_j}{x}\right)^{\frac{1}{\alpha}}\right| \mathcal{A}\right)= \nonumber\\
&=\frac{\pi \rho_m}{c}x^{-\frac{2}{\alpha}}E[P_j^{\frac{2}{\alpha}}]\nonumber \\
&\;\;\;\;\;\;\;\;\;\;+ \int_{ x\left(\frac{c}{\pi\rho_m}\right)^{\frac{\alpha}{2}}}^{\infty} \left(1-\frac{\pi \rho_m}{c}\left(\frac{P_j}{x}\right)^{\frac{2}{\alpha}}\right)f_P(P_j)dP_j\label{Eqn:MarCDFLast}
\end{align}
Since as $R\to\infty$, $Pr(\mathcal{A})\to 1$, by the symmetry between $X_i$ and $X_j$ and multiplying \eqref{Eqn:CDFUBLast} and \eqref{Eqn:MarCDFLast},
\begin{align}
&\lim_{R\to\infty}{\rm{Pr}}\left(r_i  <  \left(\frac{N^{\frac{\alpha}{2}}P_i}{x}\right)^{\frac{1}{\alpha}},\,  r_j  <  \left(\frac{N^{\frac{\alpha}{2}}P_j}{x}\right)^{\frac{1}{\alpha}}\right)\to \nonumber \\
&\lim_{R\to\infty}{\rm{Pr}}\left(  r_j  <  \left(\frac{1}{x}{N^{\frac{\alpha}{2}}P_j}\right)^{\frac{1}{\alpha}}\right){\rm{Pr}}\left(  r_i  <  \left(\frac{1}{x}{N^{\frac{\alpha}{2}}P_i}\right)^{\frac{1}{\alpha}}\right)\nonumber
\end{align}
This implies that
\begin{align}
&\lim_{R\to\infty}{\rm{Pr}}\left( N^{\frac{\alpha}{2}}P_i r_i^{-\alpha}  \leq x, N^{\frac{\alpha}{2}}P_j r_j^{-\alpha}  \leq x\right)\to \nonumber \\
&\lim_{R\to\infty}{\rm{Pr}}\left( N^{\frac{\alpha}{2}}P_i r_i^{-\alpha}  \leq x\right){\rm{Pr}}\left( N^{\frac{\alpha}{2}}P_j r_j^{-\alpha}  \leq x\right)\label{Eqn:ConvergenceJointFinal}
\end{align}
Substituting \eqref{Eqn:ConvergenceJointFinal} into \eqref{eqn32}, we have that as $R\to\infty$,
\begin{align}
\Pr\left(|H_n(x) - \bar{H}_n(x)|\geq \frac{\epsilon}{2}\right) \to 0.\label{Eqn:ConvergeToMeanEDF}
\end{align}
From the definitions of $H_n(x)$ and $\bar{H}_n(x)$,
\begin{align}
H_n(x) &= \frac{1}{n}\sum_{i = 1}^n\Ind{N^{\frac{\alpha}{2}}P_i r_i^{-\alpha} \leq x}\\
\bar{H}_n(x) &= E\left[\frac{1}{n}\sum_{i = 1}^n\Ind{N^{\frac{\alpha}{2}}P_i r_i^{-\alpha} \leq x}\right] \\
&= \frac{1}{n}\sum_{i = 1}^n\Pr\left(N^{\frac{\alpha}{2}}P_i r_i^{-\alpha} \leq x \right)
\end{align}
Substituting \eqref{Eqn:MarCDFLast} and recalling that $\Pr(\mathcal{A})\to 1$, we have
\begin{align}
&\lim_{R\to\infty}\bar{H}_n(x) =  1-\frac{\rho_m\, x^{-\frac{2}{\alpha}}}{c}E[P_j^{\frac{2}{\alpha}}]\nonumber \\
&\;\;\;\;\;\;-\int_{ x\left(\frac{c}{\pi\rho_m}\right)^{\frac{\alpha}{2}}}^{P_M} \left(1-\frac{\rho_m\left(\frac{P_j}{x}\right)^{\frac{2}{\alpha}}}{c}\right)f_P(P_j)dP_j
\end{align}
The previous expression implies that for each $\epsilon$, there exists an $R_0$, such that $\forall R > R_0$
\begin{align*}
&\left|\bar{H}_n(x) -  \left[1-\frac{\rho_m\, x^{-\frac{2}{\alpha}}}{c}E[P_j^{\frac{2}{\alpha}}]\right.\right. \nonumber \\
&\left.\left.-\int_{ x\left(\frac{c}{\pi\rho_m}\right)^{\frac{\alpha}{2}}}^{P_M} \left(1-\frac{\rho_m\left(\frac{P_j}{x}\right)^{\frac{2}{\alpha}}}{c}\right)f_P(P_j)dP_j\right]\right| < \frac{\epsilon}{2}.
\end{align*}
 When combined with \eqref{Eqn:ConvergeToMeanEDF}, we have the desired result:
\begin{align}
&\Pr\left(\left|H_n(x) -  \left[1-\frac{\rho_m x^{-\frac{2}{\alpha}}}{c}E[P_j^{\frac{2}{\alpha}}]\right.\right.\right. -\nonumber \\
&\left.\left.\left.\int_{ x\left(\frac{c}{\pi\rho_m}\right)^{\frac{\alpha}{2}}}^{P_M} \!\left(1-\frac{\rho_m\left(\frac{P_j}{x}\right)^{\frac{2}{\alpha}}}{c}\right)f_P(P_j)dP_j\right]\right| < \epsilon\right) \to 0\,.\nonumber
\end{align}
%In other words $H_n(x)$ converges in probability to the term on the RHS of the absolute value in the equation above. 

\subsection{Minimum Eigenvalue Condition}\label{Sec:MinEigenvalueProof}

Letting $\mathcal{T}$ be the set of mobiles whose transmit power is non-zero we have:
\begin{align}
&\frac1N\mathbf{S \Psi S}^\dagger % \frac1N \sum_{i = 1}^nN^{\alpha/2}R^{-\alpha} P_i \mathbf{g}_i\mathbf{g}_i^\dagger \nonumber \\
%&\;\;\;\;\;\;\;\;\;\;\;\;\;\;\;\;\;\;\;\;\;\;\;\;\;\;\;\;+ \frac1N \sum_{i = 1}^nN^{\alpha/2}(r_i^{-\alpha}-R^{-\alpha}) P_i \mathbf{g}_i\mathbf{g}_i^\dagger\nonumber \\
%= \left(\frac{c}{\pi \rho_p}\right)^{\frac{\alpha}{2}}\frac1N \sum_{i = 1}^n P_i \mathbf{g}_i\mathbf{g}_i^\dagger + \frac1N \sum_{i = 1}^nN^{\alpha/2}(r_i^{-\alpha}-R^{-\alpha}) P_i \mathbf{g}_i\mathbf{g}_i^\dagger\,.\nonumber
%\end{align}
 = \frac{N^{\frac{\alpha}{2}-1}}{R^{\alpha}}\sum_{i \in \mathcal{T}} P_i\mathbf{g}_i\mathbf{g}_i^\dagger \nonumber \\
&\;\;\;\;\;\;\;\;\;\;\;\;\;\;\;+ N^{\frac{\alpha}{2}-1} \sum_{i \in\mathcal{T}}P_i (r_i^{-\alpha}-R^{-\alpha})  \mathbf{g}_i\mathbf{g}_i^\dagger\label{Eqn:MinEvalLemmaFromJournal}\\
& =P_{\ell b} \left[\frac{N^{\frac{\alpha}{2}-1}}{R^{\alpha}}\sum_{i \in \mathcal{T}} \mathbf{g}_i\mathbf{g}_i^\dagger + N^{\frac{\alpha}{2}-1} \sum_{i \in\mathcal{T}} (r_i^{-\alpha}-R^{-\alpha})  \mathbf{g}_i\mathbf{g}_i^\dagger\right]\nonumber \\
&+ \frac{N^{\frac{\alpha}{2}-1}}{R^{\alpha}}\sum_{i \in \mathcal{T}} (P_i-P_{\ell b})\mathbf{g}_i\mathbf{g}_i^\dagger \nonumber \\
&\;\;\;\;\;\;\;\;\;\;\;\;\;\;+ N^{\frac{\alpha}{2}-1} \sum_{i \in\mathcal{T}}(P_i-P_{\ell b}) (r_i^{-\alpha}-R^{-\alpha})  \mathbf{g}_i\mathbf{g}_i^\dagger,\label{Eqn:LBLemmaProof}
\end{align}
where \eqref{Eqn:MinEvalLemmaFromJournal} follows from (39) in  \cite{CorrTransPaper}. The smallest eigenvalue of the matrix in the brackets in \eqref{Eqn:LBLemmaProof} was shown to be bounded from below by a positive value $\tilde{\lambda}_{\ell b}$ with probability 1 for $N$ sufficiently large, in Lemma 3 of  \cite{CorrTransPaper}. Moreover, since $P_i \geq P_{\ell b}$ for all $i\in \mathcal{T}$, the remaining two matrices in the sum on the RHS of  \eqref{Eqn:LBLemmaProof} are non-negative definite. Thus, by the Weyl inequality, the smallest eigenvalue of the matrix on the LHS of \eqref{Eqn:LBLemmaProof} is bounded from below by $\lambda_{\ell b} = P_{\ell b} \tilde{\lambda}_{\ell b}$, with probability 1 for $N$ sufficiently large.

\bibliography{IEEEabrv,main}

\newcommand{\noopsort}[1]{} \newcommand{\printfirst}[2]{#1}
  \newcommand{\singleletter}[1]{#1} \newcommand{\switchargs}[2]{#2#1}
\begin{thebibliography}{10}
\providecommand{\url}[1]{#1}
\csname url@rmstyle\endcsname
\providecommand{\newblock}{\relax}
\providecommand{\bibinfo}[2]{#2}
\providecommand\BIBentrySTDinterwordspacing{\spaceskip=0pt\relax}
\providecommand\BIBentryALTinterwordstretchfactor{4}
\providecommand\BIBentryALTinterwordspacing{\spaceskip=\fontdimen2\font plus
\BIBentryALTinterwordstretchfactor\fontdimen3\font minus
  \fontdimen4\font\relax}
\providecommand\BIBforeignlanguage[2]{{%
\expandafter\ifx\csname l@#1\endcsname\relax
\typeout{** WARNING: IEEEtran.bst: No hyphenation pattern has been}%
\typeout{** loaded for the language `#1'. Using the pattern for}%
\typeout{** the default language instead.}%
\else
\language=\csname l@#1\endcsname
\fi
#2}}

\bibitem{marzetta2010noncooperative}
T.~L. Marzetta, ``Noncooperative cellular wireless with unlimited numbers of
  base station antennas,'' \emph{{IEEE} Trans. Wireless Commun.}, vol.~9,
  no.~11, pp. 3590--3600, Nov. 2010.

\bibitem{LarssonMag}
E.~G. Larsson, F.~Tufvesson, O.~Edfors, and T.~L. Marzetta, ``Massive {MIMO}
  for next generation wireless systems,'' \emph{IEEE Commun. Mag.}, vol.~52,
  no.~2, pp. 186--195, Feb. 2014.

\bibitem{hoydis2013massive}
J.~Hoydis, S.~ten Brink, and M.~Debbah, ``Massive {MIMO} in the {UL/DL} of
  cellular networks: How many antennas do we need?'' \emph{{IEEE} J. Sel. Areas
  Commun.}, vol.~31, no.~2, pp. 160--171, Feb. 2013.

\bibitem{YatesMMSEMassive}
N.~Krishnan, R.~D. Yates, and N.~B. Mandayam, ``Cellular systems with many
  antennas: Large system analysis under pilot contamination,'' \emph{Proc.
  Allerton Conf.}, 2012.

\bibitem{wangsinr}
B.~Wang, Y.~Chang, and D.~Yang, ``On the {SINR} in massive {MIMO} networks with
  {MMSE} receivers,'' \emph{IEEE Commun. Lett.}, Sept. 2014.

\bibitem{novlan2012analytical}
T.~D. Novlan, H.~S. Dhillon, and J.~G. Andrews, ``Analytical modeling of uplink
  cellular networks,'' \emph{IEEE Trans. Wireless Commun.}, vol.~12, no.~6, pp.
  2669--2679, Jun. 2013.

\bibitem{rao2007reverse}
A.~M. Rao, ``Reverse link power control for managing inter-cell interference in
  orthogonal multiple access systems,'' \emph{IEEE VTC, Fall}, pp. 1837--1841,
  2007.

\bibitem{coupechoux2011set}
M.~Coupechoux and J.-M. Kelif, ``How to set the fractional power control
  compensation factor in {LTE}?'' \emph{IEEE Sarnoff Symp.}, 2011.

\bibitem{CellularNetworks}
S.~Govindasamy, D.~Bliss, and D.~H. Staelin, ``Asymptotic spectral efficiency
  of the uplink in spatially distributed wireless networks with multi-antenna
  base stations,'' \emph{{IEEE} Trans. Commun.}, vol.~61, no.~7, pp. 100--112,
  Jun. 2013.

\bibitem{govindasamy2014uplink}
S.~Govindasamy, ``Uplink performance of large optimum-combining antenna arrays
  in {P}oisson-cell networks,'' \emph{Proc. IEEE ICC.}, 2014.

\bibitem{bliss2013adaptive}
D.~W. Bliss and S.~Govindasamy, \emph{Adaptive Wireless Communications: MIMO
  Channels and Networks}.\hskip 1em plus 0.5em minus 0.4em\relax Cambridge
  University Press, 2013.

\bibitem{TxCSIJournal}
S.~Govindasamy, D.~W. Bliss, and D.~H. Staelin, ``Asymptotic spectral
  efficiency of multi-antenna links in wireless networks with limited {T}x
  {CSI},'' \emph{{IEEE} Trans. Inf. Theory}, vol.~58, no.~8, pp. 5375--5387,
  2012.

\bibitem{CorrTransPaper}
S.~Govindasamy, ``Asymptotic data rates of receive diversity systems with
  {MMSE} estimation and spatially correlated interferers,'' \emph{IEEE Trans.
  Commun.}, vol.~62, no.~5, pp. 100--113, May 2014.

\bibitem{Mathai}
A.~M. Mathai, \emph{An Introduction to Geometrical Probability}.\hskip 1em plus
  0.5em minus 0.4em\relax Gordon and Breach Science Publishers, 1999.

\end{thebibliography}

\end{document}